\documentclass[super,sort&compress]{ws-ijmpb}

\usepackage{xcolor}
\usepackage{multirow}
\usepackage{cite}
\usepackage[verbose,hypertexnames=false]{hyperref}
\hypersetup{colorlinks=false,allbordercolors=blue,pdfborderstyle={/S/U/W 1}}
\newcommand{\dd}{\mathrm{d}}

\makeatletter
\renewcommand{\@cite}[1]{\textsuperscript{#1}}
\makeatother
\begin{document}

\markboth{Z. C. Tu}
{Axisymmetric membrane shapes in the Willmore and Helfrich models}

%%%%%%%%%%%%%%%%%%%%% Publisher's Area please ignore %%%%%%%%%%%%%%%
%
\catchline{}{}{}{}{}
%
%%%%%%%%%%%%%%%%%%%%%%%%%%%%%%%%%%%%%%%%%%%%%%%%%%%%%%%%%%%%%%%%%%%%

\title{Axisymmetric membrane shapes in the Willmore and Helfrich models:
first integrals and a hyperbolic formulation}

\author{Z. C. Tu}

\address{School of Physics and Astronomy, Beijing Normal University\\
Beijing 100875, China\\
tuzc@bnu.edu.cn}

\maketitle

\begin{abstract}
The equilibrium shapes of fluid lipid membranes are governed by
nonlinear differential equations derived from the Helfrich bending
energy. In the tensionless, pressure-free, and zero-spontaneous-
curvature limit, the governing equation reduces to the
Willmore equation. We show that the combination of the Zheng--Liu and
Langer--Singer first integrals reduces the third-order axisymmetric Willmore equation to a first-order ordinary differential equation. This formulation recovers the sphere, axisymmetric minimal surfaces, and the Clifford torus as special cases, while organizing the local solution space according to two integration constants. For nonzero spontaneous
curvature, the axisymmetric Helfrich functional is reformulated
as the energy of an inhomogeneous hyperbolic elastica with
a preferred curvature proportional to the distance from the
axis. This formulation identifies
a constant-mean-curvature branch and a logarithmic branch associated
with a biconcave disk profile. The resulting exact relations
provide analytical constraints and benchmark solutions for configurations of axisymmetric vesicles and biomembranes.
\end{abstract}


\section{Introduction}

The equilibrium morphology of fluid lipid membranes is governed by
the competition among bending elasticity, spontaneous curvature,
surface tension, and the pressure difference across the membrane.
Because the thickness of a lipid bilayer is much smaller than its
lateral dimensions, the membrane can be modeled at mesoscopic scales
as a two-dimensional fluid surface embedded in three-dimensional
space. Within this continuum description, curvature elasticity
provides a unified framework for investigating vesicles, tubular and
toroidal membranes, biconcave profiles, and other biomembrane
morphologies~\cite{Seifert97,Ou-Yangbook,Lipowsky2020,Deserno2015,TuOY2014}.

The fundamental curvature energy of a fluid lipid membrane is the
Helfrich functional~\cite{Helfrich1973},
\begin{equation}\label{eq-HelfrichFun}
F_H=\frac{k_c}{2}\int_M (2H+c_0)^2\,\dd A,
\end{equation}
where $M$ denotes the membrane surface, $H$ is the mean curvature,
$\dd A$ is the area element, $k_c$ is the bending rigidity, and
$c_0$ is the spontaneous curvature. The spontaneous-curvature term
accounts phenomenologically for an asymmetry between the two sides
of the membrane. When the total area and enclosed volume are
constrained, Zhong-can and Helfrich obtained the equilibrium shape
equation~\cite{OYPRL87,OuYangHelfrich1989},
\begin{equation}\label{eq-ZCH}
k_c\nabla^2(2H)
+k_c(2H+c_0)(2H^2-2K-c_0H)
-2\lambda H+p=0,
\end{equation}
where $K$ is the Gaussian curvature, $\lambda$ is the Lagrange
multiplier associated with the area constraint, and $p$ denotes the
pressure difference across the membrane.

The Zhong-can--Helfrich equation (\ref{eq-ZCH}) has been widely used to describe
closed and open lipid vesicles, toroidal configurations, biconcave
profiles, and membrane shapes with free edges
\cite{OYPRA90,NaitoPRE1993,NaitoPRL4345,MladenovaEPJB,Mladenov2008,TuJGSP,
CapovillaPRE02,GuvenJPA02,TuJPA04,r5zhouIJMPB10,WuOuyang2025,OYXT2026}.
Within the same curvature framework, perturbative and numerical
studies have produced biconcave, peanut-like, myelin-like, and
triconcave vesicle morphologies
\cite{ZhouIJMPB2001,ZhangIJMPB2002}.
These studies illustrate the physical richness of the Helfrich model,
but they also reflect the mathematical difficulty of its nonlinear
fourth-order shape equation. Exact solutions and first-integral
reductions are therefore valuable not only for understanding special
equilibrium morphologies, but also as analytical constraints and
benchmark solutions for numerical shooting, continuation, and
surface-evolution methods.

An important limiting case is obtained by setting vanishing
$c_0$, $\lambda$ and $p$. The Helfrich functional then reduces to the
quadratic bending energy
\begin{equation}\label{eq-Poisson}
F_P=\frac{k_c}{2}\int_M (2H)^2\,\dd A,
\end{equation}
which was introduced in shell theory by Poisson
\cite{ch3-Poisson} and later employed by Canham to investigate the
biconcave shape of red blood cells~\cite{ch3-Canham70}. Its
Euler--Lagrange equation is the Willmore equation,
\begin{equation}\label{eq-Willmoreeq}
\nabla^2H+2H(H^2-K)=0.
\end{equation}
The corresponding curvature functional is also central in
differential geometry. In particular, Willmore conjectured that the
Clifford torus, whose major-to-minor radius ratio is $\sqrt{2}$,
minimizes the bending energy among embedded tori
\cite{ch3-Willmore}. This conjecture was subsequently proved by
Marques and Neves~\cite{ch3-MarquesAM}. From the membrane-physics
viewpoint, the same Clifford torus represents a distinguished
toroidal equilibrium geometry in the tensionless and pressure-free
limit.

In a general coordinate representation, the
Zhong-can--Helfrich equation is a fourth-order nonlinear partial
differential equation. Under the assumption of axial symmetry,
Hu and Ou-Yang transformed it into a third-order ordinary
differential equation~\cite{ch4-HuOY}. Zheng and Liu subsequently
reduced this equation to second order by deriving a first integral
\cite{ZhengLiu1993}. The same conserved quantity can also be
understood from the Hamiltonian and stress-tensor viewpoints
\cite{GuvenJPA05,GuvenJPA07}. In the special case of vanishing $p$, $\lambda$, $c_0$, and
integration constant of the Zheng--Liu first integral, the reduced equation is integrable and can be
transformed into a first-order ODE
\cite{Vassilev,ZhangMcDarghTu2018}. A natural question is whether
a useful first-order relation remains available when the
integration constant of the Zheng--Liu first integral is nonzero.

A complementary geometric description follows from the work of
Langer and Singer~\cite{LangerSinger1984,LangerSinger1984b}, who
showed that the meridian of a Willmore surface of revolution can be
interpreted as an elastic curve in the hyperbolic plane. This
correspondence was developed further by Bryant and Griffiths
\cite{Bryant1986} within a variational-reduction framework. The
central quantity is the signed geodesic curvature of the associated
hyperbolic elastic curve, parametrized by hyperbolic arclength. Its
equation of motion possesses an energy-type first integral, referred
to below as the Langer--Singer first integral. Hyperbolic elastica
methods have subsequently been used to study explicit Willmore
surfaces of revolution, boundary-value problems, symmetry breaking,
and the Willmore flow of tori of revolution
\cite{DallAcqua2008,DallAcqua2017,Mandel2018}.

In the present paper, we combine the Zheng--Liu and Langer--Singer
first integrals to obtain a closed first-order relation for the
axisymmetric Willmore equation that yields a branch-wise
first-order reduction of the original third-order equation. The two
integration constants in the Zheng--Liu and Langer--Singer
first integrals provide a natural way of
classifying the admissible local branches, while the spherical,
minimal-surface, and Clifford-torus solutions are recovered as
representative exact cases.
We also examine how this geometric structure is modified by a
nonzero spontaneous curvature. In the tensionless and pressure-free
sector, the axisymmetric Helfrich functional can be
reformulated as the energy of an inhomogeneous hyperbolic elastica
whose preferred curvature is proportional to the distance from the
axis. This representation clarifies why the
Langer--Singer integral is lost when $c_0\neq 0$, and it
identifies a constant-mean-curvature branch together with a
logarithmic branch associated with a biconcave, disk-like membrane
profile. These analytical relations may serve as useful constraints
and benchmark profiles in studies of axisymmetric vesicles
and biomembranes.

The remainder of the paper is organized as follows. In
section~\ref{sec-shapeeq}, we present the axisymmetric
Zhong-can--Helfrich equation and its Willmore limit. In
section~\ref{sec-firstint}, we revisit the Zheng--Liu and
Langer--Singer first integrals and combine them into a first-order relation. In section~\ref{sec-examples}, we examine the
sphere and the Clifford torus as representative exact geometries. In
section~\ref{sec-classif}, we make a classification of solutions to axisymmetric Willmore equations. In section~\ref{sec-Helfricheq}, we formulate
the axisymmetric Helfrich energy in terms of an inhomogeneous
hyperbolic elastica and discuss its constant-mean-curvature and
biconcave branches. The main results are summarized in the final
section.

\section{Axisymmetric shape equation\label{sec-shapeeq}}

\begin{figure}[!htp]
  \centering
  \includegraphics[width=7cm]{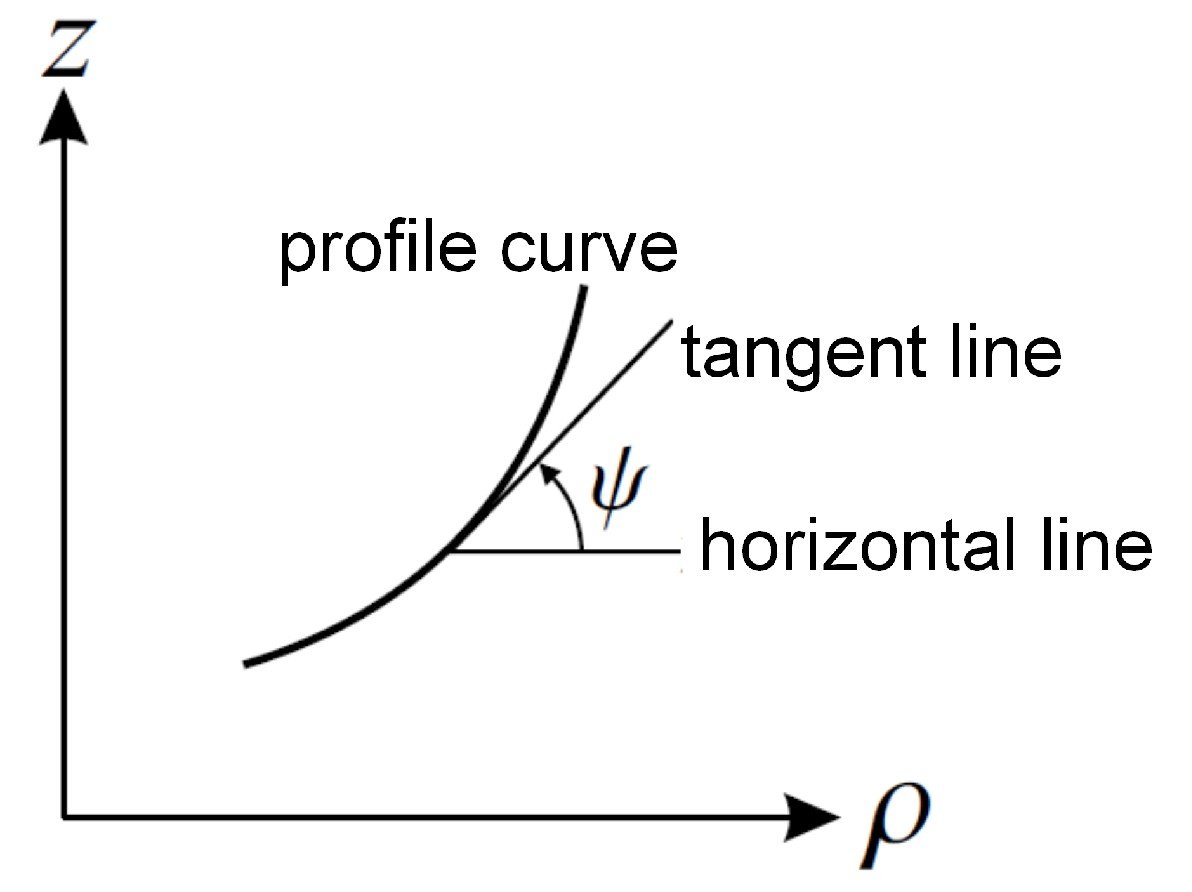}
  \caption{Profile of the generated curve.}\label{fig-profile}
\end{figure}

Consider an axisymmetric surface generated by rotating a planar profile curve $\gamma$ shown in Fig.~\ref{fig-profile} around the $z$-axis. The surface can be parametrized as
\begin{eqnarray}
x&=&\rho\cos\phi  \\
y &=&\rho\sin\phi \\
z &=& z_0+\int^\rho_{\rho_0} \tan\psi(\rho)\,\dd\rho
\end{eqnarray}
where $\psi$ is the angle between the tangent to the profile curve and the horizontal direction.
Then the mean curvature and the Gaussian curvature can be expressed as
\begin{equation}\label{eq-Haxisym}
H=-(\rho\sin\psi)^{\prime}/2\rho,
\end{equation}
\begin{equation}\label{eq-Kaxisym}
K=(\sin^{2}\psi)^{\prime}/2\rho,
\end{equation}
respectively. Throughout this paper, a prime denotes differentiation with respect to $\rho$.
For an axisymmetric scalar function, the Laplace--Beltrami operator is
\begin{equation}\label{eq-Laplaceaxisym}
\nabla^{2}=\frac
{1}{\rho^{2}}\frac{\partial^{2}}{\partial\phi^{2}}+\frac{\cos\psi}{\rho
}\frac{\partial}{\partial\rho}\left(\rho\cos\psi\frac{\partial}{\partial
\rho}\right).
\end{equation}

Substituting Eqs.~(\ref{eq-Haxisym})--(\ref{eq-Laplaceaxisym}) into the Zhong-can--Helfrich equation (\ref{eq-ZCH}), we obtain the shape equation for axisymmetric lipid vesicles~\cite{ch4-HuOY}:
\begin{align}
&-\frac{\cos\psi}{\rho}\left\{  \rho\cos
\psi\left[\frac{\left(  \rho\sin\psi\right)
^{\prime}}{\rho}\right]^\prime\right\}^\prime  -\frac{1}{2}\left[  \frac{\left(  \rho\sin
\psi\right)  ^{\prime}}{\rho}\right]  ^{3}\nonumber\\
&+\frac{\left(  \rho\sin\psi\right)  ^{\prime}\left(  \sin
^{2}\psi\right)  ^{\prime}}{\rho^{2}}-\frac{c_{0}\left(  \sin^{2}\psi\right)^{\prime}}{\rho}
+\left(\frac{\lambda}{k_{c}}+\frac{c_{0}^{2}}{2}\right)\frac{\left(\rho\sin\psi\right)  ^{\prime}}{\rho
}+\frac{p}{k_c}=0.\label{eq4-shapeaxisym}
\end{align}
If $c_0=0$, $\lambda=0$, and $p=0$, this equation reduces to the axisymmetric Willmore equation:
\begin{align}
-\frac{\cos\psi}{\rho}\left\{  \rho\cos
\psi\left[\frac{\left(  \rho\sin\psi\right)
^{\prime}}{\rho}\right]^\prime\right\}^\prime  -\frac{1}{2}\left[  \frac{\left(  \rho\sin
\psi\right)  ^{\prime}}{\rho}\right]  ^{3}
 +\frac{\left(  \rho\sin\psi\right)  ^{\prime}\left(  \sin
^{2}\psi\right)  ^{\prime}}{\rho^{2}}=0.\label{eq4-Willaxisym}
\end{align}

\section{First integrals\label{sec-firstint}}
In this section, we will revisit the Zheng--Liu and Langer--Singer first integrals and combine them into a first-order ODE.

\subsection{Zheng--Liu first integral}

As pointed out by Zheng and Liu, Eq.~(\ref{eq4-shapeaxisym}) has a first integral~\cite{ZhengLiu1993}, which can be expressed as
\begin{equation}
\frac{\Psi^{3}-\Psi( \rho \Psi^{\prime})^{2}}{2\rho
}-\rho(1-\Psi^{2})  \left[  \frac{(  \rho\Psi)
^{\prime}}{\rho}\right]  ^{\prime}-c_{0} \Psi^{2}+\left(\frac{\lambda}{k_{c}}+\frac{c_{0}^{2}}{2}\right)\rho\Psi  +\frac{p\rho^{2}}{2 k_{c}}=C_1,\label{eq4-1stIntg}
\end{equation}
where $C_1$ is a constant of integration. Note that in the Zheng--Liu first integral (\ref{eq4-1stIntg}), a new notation
\begin{equation}\Psi\equiv \sin\psi\end{equation} has been introduced for simplicity. The constant $C_1$ is related to the Noether charge associated with axial symmetry~\cite{GuvenJPA07}.

Putting $c_0=0$, $\lambda=0$, and $p=0$ in Eq.~(\ref{eq4-1stIntg}), we obtain the first integral for the axisymmetric Willmore equation:
\begin{equation}
\frac{\Psi^{3}-\Psi( \rho \Psi^{\prime})^{2}}{2\rho
}-\rho(1-\Psi^{2})  \left[  \frac{(  \rho\Psi)
^{\prime}}{\rho}\right]  ^{\prime}=C_1,\label{eqWM-1stIntg}
\end{equation}
A direct rearrangement transforms this equation into an equivalent compact form
\begin{equation}
\frac{(1-\Psi^{2})^{3/2}}{2\rho\Psi^{\prime}}
\left[
\frac{\Psi^{2}-(\rho\Psi^{\prime})^{2}}
{\sqrt{1-\Psi^{2}}}
\right]^{\prime}
=C_1.
\label{eq-WM1stIntgcpt}
\end{equation}

If $C_1=0$, this equation implies another first integral
\begin{equation}\label{eq-ZMDIntg}
\frac{\Psi^{2}-(\rho\Psi^{\prime})^{2}}
{\sqrt{1-\Psi^{2}}}=I
\end{equation}
with a constant $I$. This result was obtained in previous work~\cite{Vassilev,ZhangMcDarghTu2018}. This raises the question of whether a second first integral remains available for the axisymmetric Willmore equation (\ref{eq4-Willaxisym}) when $C_1\neq 0$?

\subsection{Langer--Singer first integral}

The Langer--Singer correspondence~\cite{LangerSinger1984,LangerSinger1984b} between an axisymmetric Willmore surface and a hyperbolic elastic curve provides another first integral. The most relevant point can be translated into the notation used in the present work as follows.
The axisymmetric Willmore equation (\ref{eq4-Willaxisym}) can be rewritten as
\begin{equation}\label{eq-Willaxisym2}
\rho\sqrt{1-\Psi^2}[\rho^2\sqrt{1-\Psi^2}\,\Psi^{\prime\prime}]^\prime+\frac{1}{2}(\rho\Psi^\prime-\Psi)^3-(\rho\Psi^\prime-\Psi)
=0.
\end{equation}
Using $\rho\Psi^{\prime\prime}=(\rho\Psi^\prime-\Psi)^\prime$, this equation can be further rewritten as
\begin{equation}\label{eq-Willaxisym3}
\rho\sqrt{1-\Psi^2}[\rho\sqrt{1-\Psi^2}\,(\rho\Psi^\prime-\Psi)^\prime]^\prime+\frac{1}{2}(\rho\Psi^\prime-\Psi)^3-(\rho\Psi^\prime-\Psi)
=0.
\end{equation}
Equivalently, introducing the differential operator
\begin{equation}
D\equiv \rho\sqrt{1-\Psi^2}\,\frac{\dd}{\dd\rho},
\end{equation}
and the signed hyperbolic curvature 
\begin{equation}q=\rho\Psi'-\Psi,\label{eq-hyper-curv}\end{equation}
Eq.~(\ref{eq-Willaxisym3}) is transformed into the form
\begin{equation}
D^2 q+\frac{1}{2}q^3-q=0.\label{eq-Willaxisym-comp}
\end{equation}
Multiplying the above equation by $Dq$ then we obtain an energy-type first integral:
\begin{equation}\label{eq-WMenfisrtInt2}
\frac{1}{2}(Dq)^2+\frac{1}{8}(q^2-2)^2=\frac{C_2}{4},
\end{equation}
which can be further expressed in an equivalent form:
\begin{equation}\label{eq-WMfisrtInt2}
4[\rho\sqrt{1-\Psi^2}\,(\rho\Psi^\prime-\Psi)^\prime]^2+[(\rho\Psi^\prime-\Psi)^2-2]^2=C_2,
\end{equation}
We refer to Eq.~(\ref{eq-WMenfisrtInt2}) or (\ref{eq-WMfisrtInt2}) as the Langer--Singer first integral.

\subsection{First-order form of the axisymmetric Willmore equation}

The two first integrals stem from distinct geometric origins. Specifically, the Zheng--Liu first integral corresponds to the Noether charge associated with axial symmetry, whereas the Langer--Singer first integral originates from the conformal invariance of the energy functional (\ref{eq-Poisson}). Hence, these two integrals are generally independent of each other, which naturally allows us to explore the physical consequences of their combination.

Rewriting the Zheng--Liu integral (\ref{eqWM-1stIntg}) in terms of $q$ and $Dq$ yields
\begin{equation}\label{eq-C1-Dq}
2\sqrt{1-\Psi^2}\,Dq+\Psi q^2+2q+2C_1\rho=0.
\end{equation}
Eliminating $Dq$ between Eqs.~(\ref{eq-WMfisrtInt2}) and (\ref{eq-C1-Dq}) casts the result into a sum of perfect squares:
\begin{equation}\label{eq-1stODE-withq}
\left[\frac{\Psi q^2+2q+2C_1\rho}{\sqrt{1-\Psi^2}}\right]^2
+(q^2-2)^2=C_2.
\end{equation}
Substituting $q = \rho\Psi^\prime-\Psi$ into this relation yields the first-order ODE governing axisymmetric Willmore surfaces:
\begin{equation}\label{eq-1stODE}
\left[\frac{\Psi(\rho\Psi^\prime-\Psi)^2+2(\rho\Psi^\prime-\Psi)+2C_1\rho}{\sqrt{1-\Psi^2}}\right]^2+[(\rho\Psi^\prime-\Psi)^2-2]^2=C_2.
\end{equation}
Equation (\ref{eq-1stODE}) should be understood as an implicit first-order ordinary differential equation. It is not an algebraic equation for $\Psi$ itself, but it is algebraic, in fact quartic, in the derivative variable $\rho\Psi'$. 
Expanding Eq.~(\ref{eq-1stODE}) yields the following quartic polynomial equation:
\begin{equation}\label{eq-1stODE2}
(\rho\Psi^\prime)^4+\alpha_2 (\rho\Psi^\prime)^2 + \alpha_1 (\rho\Psi^\prime)+\alpha_0 =0,
\end{equation}
where the variable coefficients are given by
\begin{align}
  \alpha_2 &= 4C_1\rho\Psi-2\Psi^2, \label{eq-alpha}\\
  \alpha_1 &= 8C_1\rho(1-\Psi^2),\label{eq-beta} \\
  \alpha_0 &= \Psi^4-4C_1\rho(\Psi^3+\Psi-C_1\rho)-(C_2-4)(1-\Psi^2)\label{eq-gamma}.
\end{align}

Along any admissible real-valued root branch of this quartic equation, we can formally express the relation as
\begin{equation}\label{eq-solu-quarteq}
\rho\Psi^\prime = G(\rho,\Psi; C_1,C_2).
\end{equation}
Subsequent integration of this first-order ODE yields the meridian shape profile $\Psi=\Psi(\rho;C_1,C_2)$, where the choice of the root branch must preserve continuity and remain strictly bounded by $|\Psi|\leq1$. 
For a nonzero $C_1$, the profile height $z$ can be explicitly reconstructed by integrating Eq.~(\ref{eq-WM1stIntgcpt}) once more~\cite{GuvenJPA05}:
\begin{equation}
z=C_0+\frac{\rho\Psi}{\sqrt{1-\Psi^2}}-\frac{\Psi^2-G^2}
{2C_1\sqrt{1-\Psi^2}},
\label{eq-zFirstInt}
\end{equation}
where the arguments $(\rho; C_1, C_2)$ of $\Psi$ and $G$ are omitted for brevity.

From the membrane-physics viewpoint, the first-order relation
provides a reduced constraint on admissible axisymmetric profiles.
It can be used to screen real-valued solution branches, to guide
numerical shooting or continuation schemes, and to provide
analytical checks for simulations of bending-dominated membrane
shapes.

\section{Representative membrane geometries\label{sec-examples}}

\subsection{Spherical surface}

Consider a spherical surface with radius $R$ sketched in Fig.~\ref{fig-four-profile}(a), which can be expressed as $\Psi=\rho/R$.
Eqs.~(\ref{eqWM-1stIntg}) and (\ref{eq-WMfisrtInt2}) are simultaneously satisfied with $C_1=0$ and $C_2=4$ when $\Psi=\rho/R$.

Now we check the consistency of Eq.~(\ref{eq-1stODE2}). If $C_1=0$ and $C_2=4$, we have $\alpha_2=-2\Psi^2$, $\alpha_1=0$, and $\alpha_0=\Psi^4$ from Eqs.~(\ref{eq-alpha})--(\ref{eq-gamma}). Then Eq.~(\ref{eq-1stODE2}) reduces to $(\rho\Psi^\prime)^4-2\Psi^2 (\rho\Psi^\prime)^2 +\Psi^4 =0$, which implies $\rho\Psi^\prime=\pm\Psi$. The solution $\Psi=\rho/R$ corresponds to the positive branch $\rho\Psi^\prime=\Psi$. The negative branch corresponds to an axisymmetric minimal surface, $\Psi=A/\rho$, where $A$ is a constant.

\subsection{Clifford torus}

\begin{figure}[!htp]
  \centering
  \includegraphics[width=7cm]{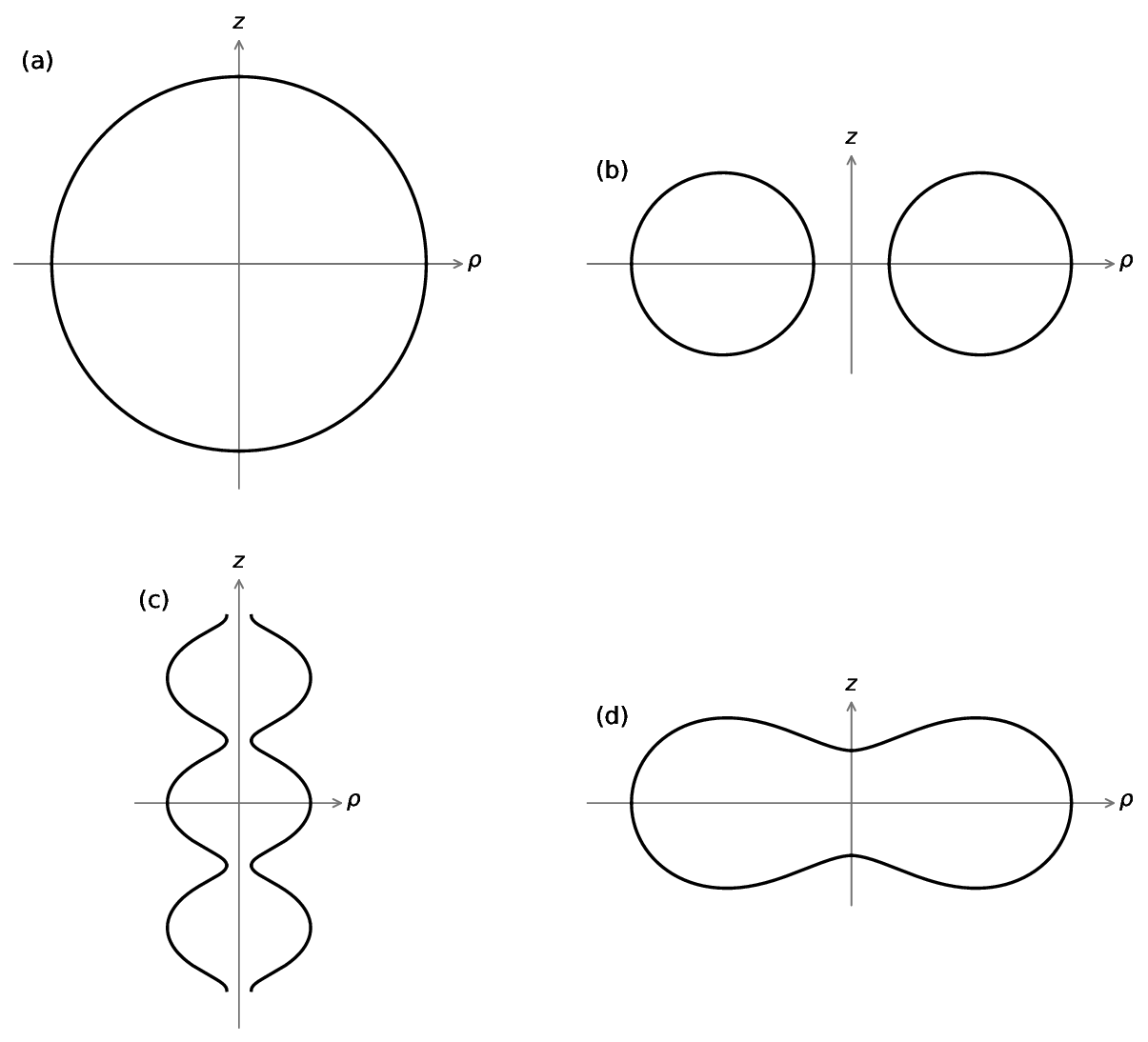}
  \caption{Schematic plot of representative axisymmetric membrane profiles: (a) spherical surface, (b) Clifford-torus, (c) constant-mean-curvature surface, and (d) biconcave disk.}\label{fig-four-profile}
\end{figure}

The Clifford torus sketched in Fig.~\ref{fig-four-profile}(b) is a classical Willmore surface. It may be represented as a torus of revolution whose generating circle has major radius $R$ and minor radius $r$ with ratio
\begin{equation}\label{eq-Clifford}
\frac{R}{r}=\sqrt{2}.
\end{equation}
This torus can be expressed as $\Psi=\rho/r-\sqrt{2}$. In this case,
Eqs.~(\ref{eqWM-1stIntg}) and (\ref{eq-WMfisrtInt2}) are simultaneously satisfied with $C_1=-1/r$ and $C_2=0$.

Conversely, since Eq.~(\ref{eq-1stODE}) is a sum of perfect squares, we can derive
\begin{eqnarray}
% \nonumber % Remove numbering (before each equation)
\Psi(\rho\Psi^\prime-\Psi)^2+2(\rho\Psi^\prime-\Psi)-2\rho/r &=& 0 \\
(\rho\Psi^\prime-\Psi)^2-2&=& 0
\end{eqnarray}
when $C_1=-1/r$ and $C_2=0$. The above equations lead to $\Psi=\rho/r-\sqrt{2}$ (the other solution $\Psi=\rho/r+\sqrt{2}$ is omitted since it exceeds 1).

\section{Classification of solutions to axisymmetric Willmore equation\label{sec-classif}}
Because $C_1$ enters Eq.~(\ref{eq-1stODE}) only through the product $C_1\rho$, every nonzero $C_1$ can be normalized to $C_1=\pm1$ by a rescaling of $\rho$. Thus, up to scale, the local branches fall into the sectors $C_1=0,+1,-1$ which correspond to zero-, positive- or negative Noether charge, respectively. The second constant $C_2$ is nonnegative and bounds the signed hyperbolic curvature $q$.
Indeed, Eq.~(\ref{eq-1stODE-withq}) implies
\begin{equation}
\max\!\left(0,2-\sqrt{C_2}\right)\leq q^2\leq2+\sqrt{C_2}.
\end{equation}
Consequently, $C_2=0$ forces $q=\pm\sqrt{2}$; for $0<C_2<4$, $q$ cannot vanish; $C_2=4$ is the threshold at which $q=0$ becomes possible; and $C_2>4$ allows sign changes of $q$.

The zero-charge sector requires separate attention. Setting $C_1=0$ in Eq.~(\ref{eq-1stODE}) gives
\begin{equation}\label{eq-zeroNcharge}
(\rho^2\Psi'^2-\Psi^2)^2=(C_2-4)(1-\Psi^2).
\end{equation}
For a regular local branch with $|\Psi|<1$, Eq.~(\ref{eq-zeroNcharge}) requires $C_2\geq4$. Hence there are no regular zero-charge branches with $0\leq C_2<4$. At $C_2=4$, the elementary factors $\rho\Psi'=\pm\Psi$ yield the spherical branch $\Psi=\rho/R$ and the axisymmetric minimal-surface branch $\Psi=A/\rho$.

For $C_1=-1$ and $C_2=0$, the admissible constant-curvature choice is $q=\sqrt{2}$, giving $\Psi=\rho-\sqrt{2}$, i.e. the Clifford torus after restoring the scale. The branch $q=-\sqrt{2}$ gives $\Psi=\rho+\sqrt{2}>1$ and is inadmissible. The $C_1=+1$ sector is obtained by the corresponding orientation-reversed branch. Table~\ref{tab-classification} summarizes possible classification of solutions to axisymmetric Willmore equation.

\begin{table}[ht]
\caption{\label{tab-classification}
Classification of solutions to axisymmetric Willmore equation}
\centering
\footnotesize
\renewcommand{\arraystretch}{1.3}
\setlength{\tabcolsep}{4pt}
\begin{tabular}{|c|c|p{4.2cm}|p{6.0cm}|}
\hline
$C_1$ & $C_2$ & \textbf{Behavior of $q=\rho\Psi'-\Psi$} & \textbf{Representative information} \\
\hline
\multirow{3}{*}{$0$}
& $0\leq C_2<4$ & No regular branch with $|\Psi|<1$ & Excluded by Eq.~(\ref{eq-zeroNcharge}). \\
\cline{2-4}
& $4$ & $q$ may reach zero; $0\leq q^2\leq4$ & Includes $\Psi=\rho/R$ and $\Psi=A/\rho$. \\
\cline{2-4}
& $>4$ & $q$ may change sign & More general zero-charge Willmore branches. \\
\hline
\multirow{3}{*}{$+1$}
& $0$ & $q=-\sqrt{2}$& Orientation-reversed normalized Clifford-torus branch $\Psi=\sqrt{2}-\rho$; the opposite sign is inadmissible. \\
\cline{2-4}
& $0<C_2<4$ & $q$ has fixed nonzero sign & Positive-charge branches with fixed-sign hyperbolic curvature. \\
\cline{2-4}
& $\geq4$ & $q=0$ is possible; sign changes occur for $C_2>4$ & Critical and sign-changing positive-charge branches. \\
\hline
\multirow{3}{*}{$-1$}
& $0$ & $q=\sqrt{2}$ & Normalized Clifford-torus branch $\Psi=\rho-\sqrt{2}$; the opposite sign is inadmissible. \\
\cline{2-4}
& $0<C_2<4$ & $q$ has fixed nonzero sign & Negative-charge branches with fixed-sign hyperbolic curvature. \\
\cline{2-4}
& $\geq4$ & $q=0$ is possible; sign changes occur for $C_2>4$ & Critical and sign-changing negative-charge branches. \\
\hline
\end{tabular}
\end{table}

\section{Axisymmetric Helfrich surfaces\label{sec-Helfricheq}}
To isolate the effect of spontaneous curvature, we restrict the
following discussion to the tensionless and pressure-free sector,
$\lambda=0$ and $p=0$. In this section, we examine how the preceding hyperbolic formulation is modified when the spontaneous curvature $c_0$ is nonzero in the Helfrich functional (\ref{eq-HelfrichFun}). Here, an unconstrained stationary configuration of this functional is referred to as the Helfrich surface which is governed by the following equation:
\begin{equation}\label{eq-ZCH-reduce}
\nabla^2 (2H)+( 2H+c_0)(2H^{2}-2K-c_0H)=0.
\end{equation}

In the axisymmetric case, setting $\lambda=0$ and $p=0$ in Eq.~(\ref{eq4-1stIntg}), we have the first integral:
\begin{equation}
\frac{\Psi^{3}-\Psi( \rho \Psi^{\prime})^{2}}{2\rho
}-\rho(1-\Psi^{2})  \left[  \frac{(  \rho\Psi)
^{\prime}}{\rho}\right]  ^{\prime}-c_{0} \Psi^{2}+\frac{c_{0}^{2}\rho\Psi}{2} =C_1.\label{eq4-1stIntg-c1}
\end{equation}

\subsection{Helfrich energy as an inhomogeneous hyperbolic elastica}
Under the Langer--Singer correspondence, the meridian of an axisymmetric Willmore surface is interpreted as an elastic curve in the hyperbolic plane \cite{LangerSinger1984,LangerSinger1984b}, from which the first Langer--Singer integral (\ref{eq-WMfisrtInt2}) naturally emerges. Nevertheless, this elegant geometric picture is disrupted for axisymmetric Helfrich surfaces, as the $c_0$ term in the Helfrich functional spoils the underlying conformal symmetry. This motivates us to apply the Langer--Singer framework to further investigate the Helfrich functional.

The hyperbolic plane $\{(z,\rho)\}$ with $\rho\ge 0$, equipped with metric
\begin{equation}
 \dd s_H^2=\frac{\dd\rho^2+\dd z^2}{\rho^2}=\frac{\dd\rho^2}{(1-\Psi^2)\rho^2},
 \label{eq-hyperbolic-metric}
\end{equation}
has negative Gaussian curvature $-1$. Along the meridian, the hyperbolic arclength element is
\begin{equation}
 \dd s_H=\frac{\dd\rho}{\rho \sqrt{1-\Psi^2}}.
 \label{eq-hyperbolic-arclength}
\end{equation}
Considering the signed hyperbolic geodesic curvature (\ref{eq-hyper-curv}), we have $(\rho\Psi)'=\rho\Psi'+\Psi=q+2\Psi$. Thus the mean curvature (\ref{eq-Haxisym}) can be expressed as
\begin{equation}
H=-\frac{q+2\Psi}{2\rho}. \label{eq-Haxisym-hyper}
\end{equation}
In addition, we can calculate the area element, which reads
\begin{equation}
\dd A=\frac{\rho\dd \rho\dd \phi}{\sqrt{1-\Psi^2}}=2\pi\rho^2\,\dd s_H.
 \label{eq-area-hyperbolic}
\end{equation}

Substituting the above two equations into (\ref{eq-HelfrichFun}), we arrive at the hyperbolic representation of the axisymmetric Helfrich functional:
\begin{equation}
 F_H=\pi k_c\int_\gamma
 (q+2\Psi-c_0\rho)^2\,\dd s_H,
 \label{eq-Helfrich-hyper}
\end{equation} where $\gamma$ corresponds to the meridian.

Note that $(q+2\Psi-c_0\rho)^2-(q-c_0\rho)^2 =4\Psi(q+\Psi-c_0\rho) =4\rho\Psi(\Psi'-c_0)$, and $\rho\Psi(\Psi'-c_0)\dd s_H$ can be expressed in a total differential form as $
\rho\Psi(\Psi'-c_0)\,\dd s_H = -\dd (\sqrt{1-\Psi^2}+c_0z)$. Since $\int_\gamma \dd (\sqrt{1-\Psi^2}+c_0z)$ will enter the boundary term, the functional (\ref{eq-Helfrich-hyper}) is variationally
equivalent to 
\begin{equation}
 \tilde{F}_H = \pi k_c\int_\gamma
 (q-c_0\rho)^2\,\dd s_H,
 \label{eq-Helfrich-hyper2}
\end{equation}
which implies that the Helfrich meridian is an inhomogeneous hyperbolic elastica with spontaneous curvature
$q_0=c_0\rho$ which is proportional to $\rho$. If $c_0=0$, the above functional reduces to the Langer--Singer form. Its variation gives the axisymmetric Willmore equation (\ref{eq-Willaxisym-comp}), from which one obtains the Langer--Singer first integral (\ref{eq-WMfisrtInt2}).
This conservation law ultimately reflects the conformal invariance of the functional (\ref{eq-Poisson}). The presence of $c_0$ breaks the conformal invariance and the functional (\ref{eq-Helfrich-hyper2}) gives the axisymmetric Helfrich equation: 
\begin{equation}\label{eq-axi-helfrich-surf}
D^2q+\frac12q^3-q +2c_0\rho\Psi(q+\Psi) -\frac{c_0^2\rho^2}{2}(q+2\Psi)=0,
\end{equation}
which is equivalent to (\ref{eq4-shapeaxisym}) with vanishing $\lambda$ and $p$. The explicit $\rho$- and $\Psi$-dependent terms related to $c_0$ prevent Eq.~(\ref{eq-axi-helfrich-surf}) from having the same first integral (\ref{eq-WMfisrtInt2}) as the Willmore equation.

\subsection{Constant-mean-curvature and biconcave branches}
Although the spontaneous-curvature terms prevent the immediate construction of the Langer--Singer integral, the hyperbolic representation still identifies two simple exact branches. First, consider the pointwise zero-density condition for the functional (\ref{eq-Helfrich-hyper}) which corresponds to
\begin{equation}\label{eq-minsrface}
q+2\Psi-c_0\rho = 0.
\end{equation}
Combining this condition with Eq.~(\ref{eq-hyper-curv}) gives
\begin{equation}
 \Psi=\frac{c_0\rho}{2}+\frac{A}{\rho},
 \label{eq-CMC-branch}
\end{equation}
where $A$ is a constant. This branch has constant mean curvature $H=-c_0/2$ sketched in Fig.~\ref{fig-four-profile}(c).

Second, consider the pointwise zero-density condition for the variationally equivalent functional (\ref{eq-Helfrich-hyper2}) which corresponds to
\begin{equation}\label{eq-RBC}
q-c_0\rho =0.
\end{equation}
Combining this condition with Eq.~(\ref{eq-hyper-curv}) gives
\begin{equation}
 \Psi=c_0\rho
 \ln\left(\frac{\rho}{\rho_B}\right),
 \label{eq-eq-RBCsolu}
\end{equation}
where $\rho_B>0$ is a constant. This solution possesses a biconcave disk-like profile sketched in Fig.~\ref{fig-four-profile}(d), which was previously used to explain the shapes of red blood cells ~\cite{NaitoPRE1993}.

\section{Conclusion}

In summary, we have investigated axisymmetric membrane shapes
within the Willmore and Helfrich curvature models. In the Willmore
limit, the Zheng--Liu and Langer--Singer first integrals combine
into an implicit relation that is quartic in the derivative
variable $\rho\Psi'$. Away from the exceptional set, this
relation provides a branch-wise first-order reduction and organizes
the local solutions through the constants $(C_1,C_2)$. The
sphere and the Clifford torus are recovered as representative
closed geometries, while the minimal-surface solution appears as
an additional mathematical branch.

For nonzero spontaneous curvature, the tensionless and
pressure-free Helfrich energy is reformulated as the energy of an
inhomogeneous hyperbolic elastica with preferred curvature
$q_0=c_0\rho$. This representation clarifies why the autonomous
Langer--Singer energy integral is lost and yields two exact
analytical branches: a constant-mean-curvature family and a
logarithmic biconcave profile. These results provide analytical
constraints and benchmark solutions for future studies of
axisymmetric lipid-vesicle and biomembrane morphologies.

\section*{Acknowledgements}
The author is grateful for the support from the National Natural Science Foundation of China (Grant Nos. 12475032, 11975050, and 11322543).

\section*{ORCID}
\noindent Zhanchun Tu - \url{https://orcid.org/0000-0003-2548-3785}

\end{document}